**Title:** Novel Layered Iridate $Ba_7Ir_3O_{13+\delta}$ Thin Films with Colossal Permittivity

*Ludi Miao, Yan Xin, Jinyu Liu, Huiwen Zhu, Hong Xu, Diyar Talbayev, Taras Stanislavchuk, Andrei Sirenko, Venkata Puli, and Zhiqiang Mao\**

((Optional Dedication))

Dr. L. Miao, J. Liu, Dr. H. Zhu, Dr. H. Xu, Prof. D. Talbayev, Dr. V. Puli, Prof. Z.Q. Mao
Department of Physics and Engineering Physics, Tulane University, New Orleans, Louisinana 70118, USA
E-mail: zmao@tulane.edu

Prof. Y. Xin
National High Magnetic Field Laboratory, Florida State University, Tallahassee, Florida 32310, USA

T. Stanislavchuk, Prof. A. Sirenko
Department of Physics, New Jersey Institute of Technology, Newark, New Jersey 07102, USA

Transition metal oxides (TMOs) have been a hotly pursued research areas in condensed matter physics since they show a rich variety of exotic phenomena. The examples include high-temperature superconductivity in cuprates,[1,2] colossal magnetoresistance in manganites,[3] multiferroicity in $BiFeO_3$,[4] metal-to-insulator transition in vanadium oxides,[5] *etc*. These properties are governed by the strongly correlated *d* electrons, which are normally characterized by the interplay between spin, charge, lattice, and orbital degrees of freedom, leading the properties of correlated materials to be extremely sensitive to external stimuli such as magnetic field,[6] carrier doping,[7] pressure,[8] and epitaxial strain.[9]

Recently, 5$d$ TMOs such as iridates attracted great interest due to their strong spin-orbit coupling (SOC) from heavy element Ir.[10-16] The strength of SOC is comparable to the kinetic energy and the Coulomb repulsion energy in 5$d$ TMOs, making SOC an additional degree of freedom governing the physical properties. For example, without taking SOC into consideration, layered perovskite $Sr_2IrO_4$ is expected to be a Fermi liquid metal like its 4$d$ counterpart $Sr_2RhO_4$.[17] Surprisingly, $Sr_2IrO_4$ has been experimentally demonstrated to be a SOC induced Mott insulator.[10,11] Other examples of SOC induced exotic states include a metallic chiral spin liquid state in $Pr_2Ir_2O_7$,[12,13] a semi-metallic state in $Na_3Ir_3O_8$,[14] and a heavy fermion state in $CaCu_3Ir_4O_{12}$.[18] Moreover, SOC in iridates is also theoretically predicted to create new exotic phases hardly seen in 3$d$ and 4$d$ TMOs, such as topological insulating states,[19-22] spin liquid states in the Kitaev limit[23] and unconventional superconductivity.[24] As far as we know, recent experimental studies have suggested hyperhoneycomb $\beta$-$Li_2IrO_3$ is in close proximity to the Kitaev spin liquid;[25] however, other predicted exotic phenomena have not yet been found in any known iridate systems. Therefore, exploring new iridate phases may be an effective route to discover novel functional properties of iridates.

In this communication, we report the discovery of new layered insulating iridate $Ba_7Ir_3O_{13+\delta}$ (BIO) in thin film form. We will show although BIO films can be quasi-epitaxially grown on STO(111), LAO(001) and MgO(110) substrates with the same $c$-orientation, the dielectric properties of these films are strongly substrate dependent. The BIO/STO(111) and BIO/LAO(001) films exhibit colossal permittivity (CP) ~$10^4$, while the BIO/MgO(110) films permittivity is only ~50. Such a sharp contrast in dielectric properties between these films can be attributed to their distinct microstructures and the observed CP originates from the colossal internal barrier layer capacitance (IBLC) effect at *atomically thin* domain boundaries.

Our BIO/STO(111), BIO/LAO(001) and BIO/MgO(110) films were grown using the pulsed laser deposition (PLD) method (see the experimental section for details). The crystal phase and microstructures of the films are investigated by transmission and scanning transmission electron microscopy (TEM/STEM), and high resolution x-ray diffraction (HR-XRD) (see the method section for details). The crystal phase is the same for all three films on different substrates. **Figure 1**(a)-(c) are STEM high angle annular dark field (HAADF) images projected along [010], [100], and [001] directions of the BIO film. All the atoms except for oxygen are shown as the white contrasted dots. The Ir atomic columns have higher intensity contrast than the Ba atomic columns due to the larger atomic number $Z$. The projection images along [010] and [100] directions clearly show that the BIO crystal has a layered structure, with $c \sim 25.5$ Å. The projection image along [100] shows that this layered structure consists of two distinguished "A" and "B" layers. "A" layers are arrow-like, alternately pointing toward left or right, with each "arrow" consisting of two Ir atoms in the middle and three Ba atoms with one at the arrow head and two at the arrow tail; "B" layers are chain-like, with each segment consisting of an Ir atom and four Ba atoms. Based on the number of Ba and Ir atoms in the unit cell, which is deduced from three projections, and the fact that Ir ion has a valence of 4+ or 5+, we derived the chemical formula for BIO to be $Ba_7Ir_3O_{13+\delta}$. Energy dispersive x-ray spectroscopy measurements on the BIO films have indeed shown that the Ba-to-Ir ratio is 70.1:29.9. From the projection images along three directions as well as the electron diffraction patterns along corresponding directions as shown in Figure 1(e)-(g), an orthorhombic structure can be determined for the BIO film. From the diffraction patterns, the lattice constants for the BIO film are determined to be $a = 5.960$ Å, $b$

= 10.365 Å, and $c$ = 25.535 Å. It is interesting to note that the STEM HAADF image along [001] projection (Figure 1(c)) shows a hexagonal arrangement of atoms. However, the high resolution TEM image (HRTEM) along the same direction (Figure 1 (d)), which contains atomic structure information along $c$ axis, reflects the true nature of an orthorhombic structure on $ab$ plane.

The insulating properties of BIO films have been verified by optical spectroscopy measurements and transport measurements. The results of optical measurements are presented in the supplementary information, from which the optical bandgap of BIO is estimated to be 1.3 eV at room temperature, much higher than the gaps of any other known iridate compounds (~0.1 eV for $Sr_2IrO_4$,[16] and 0.34 eV for $Na_2IrO_3$[26]). The room temperature in-plane resistivity of the BIO films is ~ 110,000 Ωcm, five orders of magnitude larger than the in-plane resistivity of $Sr_2IrO_4$.

The BIO films on STO and LAO substrates show exotic dielectric properties. We have measured the in-plane permittivity $\varepsilon'_{in}$ for BIO films, as shown in **Figure 2**(a). To maximize the measured capacitance, we used the inter digital electrode (IDE) pattern [see the inset of Figure 2(a)]. The details of measurements are discussed in the methods section below. The most remarkable feature is that $\varepsilon'_{in}$ for BIO/STO(111) and BIO/LAO(001) films stays in the ~$10^4$ order of magnitude for $f < 10^5$ Hz and ~ $10^3$ for $10^5$ Hz $< f < 10^6$ Hz. Such huge permittivity values for these two films clearly fall into the CP range (> 1000) and are even comparable with those of many well-known CP materials including (Nb + In) co-doped $TiO_2$ (CP ~ $6\times10^4$ [27]), $CaCu_3Ti_4O_{12}$ (CCTO, CP ~ $10^5$ [28]), $LuFe_2O_4$[29] and $BaTiO_3$-doped perovskites[30] (CP ~ 5000) measured under the same condition. However, as compared to other CP materials, these BIO films have relatively high dielectric loss with tan$\delta$ ~ $10^2$-$10^{-1}$ in the measured frequency range (see supplemental material). In sharp contrast, $\varepsilon'_{in}$ for BIO/MgO film is too small to be measurable even with the IDE patterns. From the system resolution, we could estimate the permittivity of BIO/MgO(110) film to be less than 50. Given that all three types of films possess the same crystal phase, the in-plane CP observed in BIO/STO(111) and BIO/LAO(001) films cannot be due to a material-intrinsic mechanism, such as electron-pinned defect-dipole effect,[27] charge ordering,[29] structurally frustrated relaxor ferroelectrics, nanoscale disorder, or Mott-variable-range-hopping.[28,31]

In order to clarify the origin of distinct dielectric properties between BIO/STO(111) (BIO/LAO(001)) and BIO/MgO(110) films, we have further investigated their crystal structures through HR-XRD measurements. These films show crystallization with the *c*-orientation, regardless of the substrate orientation, which can be clearly seen from the strong (00*l*) diffraction peaks in the 2$\theta$-$\theta$ scans, as shown in **Figure 3**(a)-(c). Such a substrate-orientation-independent *c*-oriented growth for BIO films strongly suggests the existence of multiple domains, as the crystalline symmetry of the BIO films and the substrates are incompatible. To confirm this, we performed HR-XRD $\varphi$ scans for the (0 4 18) reflections of BIO/STO(111) and BIO/LAO(001) films and the (2 0 16) reflection of BIO/MgO(110) films, as shown in Figure 3(d)-(f), respectively. The 12-fold diffraction peaks for BIO/LAO(001) indicate the existence of six domain orientations, since each orthorhombic domain contributes two peaks in the $\varphi$ scan. For BIO/STO(111) and BIO/MgO(110) films, the 6-fold diffraction peaks indicate that there are three domain orientations. We also noticed that the BIO/STO(111) films have the sharpest $\varphi$ scan peaks, whereas the BIO/MgO(110) films have the widest peaks, suggesting the domain coherence is the best for the BIO/STO(111) films, but the worst for the BIO/MgO(110) films. Such a substrate dependent domain coherency is possibly due to the distinct symmetry mismatching between BIO films and substrates.

We further characterized the domain structures of these films using TEM/STEM techniques and determined the relative domain orientations for each type of film. Since the orthorhombic structure of BIO films is characterized by the hexagonal arrangement of atoms on the basal plane as shown in Figure 1c and the lattice constant *a* of BIO films are close to $\sqrt{2}$ times the lattice parameter of cubic oxide substrates, the $[100]_{BIO}$ axis would favor to align with [110]-equivalent crystal axes of the substrates. From this perspectives, combined with the domain information revealed from the XRD $\varphi$ scans, we can speculate potential domain orientations on each substrate, as illustrated in Figure 3(g)-(I). The relative orientation angle between the domains is expected to be 60°/120° for the BIO/STO(111) and BIO/MgO(110) films (see Figure 3(g) and (I)), but either 60°/120° or 90° for the BIO/LAO(001) film (Figure 3h). These expected domain orientations have been evidenced in our TEM and STEM observations, as discussed below.

For the BIO/STO film, as seen in the inset to **Figure 4** (a), it shows an electron diffraction pattern with a 6-fold symmetry, consistent with the result of the XRD $\varphi$ scan. The domain boundaries as expected from Figure 3(g) are not visible either for the [001] projection in low magnification bright field (BF) TEM image (Figure 4(a)) nor in STEM HAADF imaging (Figure 4(c)). However, they show up in HRTEM imaging, as shown in Figure 4(b), where the two domains are rotated by 120° relative to each other. These domain boundaries are *atomically thin* (denoted by type I domain boundaries hereafter).

For BIO/LAO(001) film, in addition to type I domain boundaries, the expected domain boundary characterized by a 90° rotation (see Figure 3h) is indeed observed, as shown in Figure 4 (d)-(g). The electron diffraction pattern over the whole area of Figure 4(d) exhibits a 12-fold symmetry (inset to Figure 4(d)), indicating the existence of 6 domain orientations, in agreement with the HR-XRD $\varphi$ scan results. Figure 4(e) and (f) are STEM HAADF cross sectional images of the BIO/LAO(001) film, which show the 90° domain boundaries (type II). Type II domain boundaries show up as continuous dark lines in low magnification BF image (Figure 4(d)). The 90° domain rotation can also be seen clearly from the atomic structures on the (001) plane as shown in Figure 4(g). Like type I domain boundaries, type II domain boundaries are also atomically sharp, ~ 2-3 atoms thick. Apparently, type II domain boundary can be attributed to two hexagonal atomic arrangements on the basal plane oriented perpendicularly to each other, as shown in Figure 3(h). They start to form from at the film/substrate interface and thread through the whole film to the film surface (Figure 4(e)).

For the BIO/MgO(110) film, while it is expected to exhibit similar domain orientations as the BIO/STO film as shown in Figure 3(I), it exhibits additional defects besides type I domain boundaries. As seen in the inset to Figure 4(h), although the BIO/MgO(110) film also exhibits an electron diffraction with a 6-fold symmetry along the [001] projection, the diffraction spots show a dispersion of ~7$^o$, indicating that its domain orientation is not perfectly coherent as compared to BIO/STO(111) films. This is also exactly what we have seen in the HR-XRD $\varphi$ scan where the peaks representing three different domain orientations look much broader than the other two types of films. From the low magnification BF TEM images [Figure 4(h) and 4(I)], networks of line defects can be clearly seen and these line defects form a new type of domain boundaries (type III). The typical dislocation contrast of line defects arises from the electrons deflected strongly by the distorted atomic planes around the dislocation cores.[33] These individual dislocations thread through the whole film from the film/substrate interface to the film surface [Figure 4(J)]. The atomic structure of the projected dislocation cores is presented in Figure 4(k), where these dislocations have a Burgers vector of [100] by drawing Burgers circuit around the core. Those two white lines indicate the atomic planes across the boundary have ~7° deviation in orientation, consistent with the ~7° dispersion seen in the electron diffraction. The presence of such line defect networks should be associated with strong lattice mismatch between the MgO(110) substrate and the BIO film.

The distinct dielectric properties between the BIO/STO (BIO/LAO) and the BIO/MgO film can be well understood in terms of the distinct microstructures of these films discussed above. In the BIO films on LAO and STO substrates, domain boundaries are *atomically thin* for either type I or II. Therefore, we can expect a colossal IBLC effect due to the Maxwell-Wagner polarization at domain boundaries,[28,31,32] which should be responsible for the observed CP in these two types of films. However, for the BIO/MgO film, they exhibits additional dislocation network in addition to type I domain boundaries. Since dislocation defects are known to be electrically active in oxides,[34,35] and can form conducting leakage path[36,37] through its network, which accounts for the absence of CP in the BIO/MgO(110) films.

To further demonstrate the CP observed in BIO/STO(111) and BIO/LAO(001) films is caused by the IBLC effect at *atomically thin* domain boundaries, we measured the out-of-plane permittivity $\varepsilon'_{out}$ for the BIO/SrRuO$_3$/STO(111) heterostructure where the 10-nm-thick SrRuO$_3$ buffer layer acts as the bottom electrode (see inset in Figure 2(b)). $\varepsilon'_{out}$ is around 10, three orders of magnitude smaller than $\varepsilon'_{in}$, which is consistent with the fact that out-of-plane oriented electric field does not cross domain boundaries mostly perpendicular to the film surface, as discussed above. This provides strong evidence for the above argument that the observed CP in the BIO/STO or BIO/LAO films originates from the colossal IBLC effect. In general, the IBLC effect is especially remarkable in the low frequency range ($f < 10^3$ Hz),[38] which is in line with the rapid decrease of permittivity for BIO/STO(111) and BIO/LAO(001) films as the frequency increases. To determine if the low permittivity in BIO/MgO(110) films is due to electrical leakage through the networks of threading dislocations, we also performed I-V measurements with IDE patterns for all BIO films, as shown in Figure 2(c) and (d). Indeed, the I-V curves are non-linear for BIO/STO(111) and BIO/LAO(001) films, consistent with the fact that the domain boundaries act as tunnel barriers. For BIO/MgO(110) films, however, the I-V curve is linear (see Figure 2(d)), indicating the existence of an ohmic leakage channel. The non-linear I-V curves seen in BIO/STO and BIO/LAO films (Figure 2(c)) are not caused by heating effect since the current in these films is two-orders-of-magnitude less than that in the BIO/MgO film for a given voltage.

In conclusion, we have synthesized new layered iridate Ba$_7$Ir$_3$O$_{13+\delta}$ (BIO) thin films on STO(111), LAO(001), and MgO(110) substrates using the PLD method. All BIO films are *c*-axis orientated on all substrates and characterized by insulating behavior with a band gap of ~ 1.3 eV. The BIO/STO and BIO/LAO films show CP, while the BIO/MgO(110) film does not. Our microstructural analyses by TEM/STEM show all these films display atomically thin domain boundaries, which is responsible for the CP observed in the BIO/STO and BIO/LAO films. In BIO/MgO(001) films, however, additional networks of conducting threading dislocations are present, which prevents the IBLC effect from being observed. Our findings not only suggest a new route to seeking new CP materials through domain boundary engineering, but also unveil a new platform for investigating novel SOC physics in iridates.

**Experimental Section**

*Sample preparation*: The quasi-epitaxial thin films of BIO are grown using the PLD method with a KrF excimer laser ($\lambda$ = 248 nm). Single crystalline $SrTiO_3$(111), $LaAlO_3$(001) and MgO(110) are used as substrates. A ceramic pellet with nominal compositions of $Ba_2IrO_4$ was used as a PLD target with excessive Ir being used to compensate the loss of volatile Ir during the deposition. The films were deposited at 1080°C in the atmosphere of 150 mTorr of $O_2$ for one minute. The laser pulse has an energy density of 0.7 J/cm$^2$, with a repetition rate of 10Hz. All the films are 277 nm thick. After deposition, the films are quenched in He gas.

*Structure and optical characterization*: The structures of the BIO films were characterized by a high resolution four circle XRD and a probe aberration corrected JEM-ARM200cF at 200kV. The STEM images were taken with an probe convergence semi-angle of 11 mrad and collection angles of 90-174.6 mrad. Chemical composition were determined by field-emission SEM. We measured the real part of the optical conductivity of the BIO/STO film using variable-incidence-angle ellipsometry at room temperature.

*Electrical and dielectric measurements:* The dielectric spectra of the BIO films were investigated in an intermediate frequency range ($10^2$ Hz $< f <$ $10^6$ Hz) at the room temperature using a dielectric spectrometer. IDE patterns of sputtered gold with inter-electrode distance ~ 37 μm were used for the measurements along in-plane direction. Contributions from substrates were subtracted with a separate measurement on a bare substrate with the same IDE pattern. I-V measurements were performed at room temperature with the same IDE pattern.

Top electrodes of sputtered gold with a size of ~ 50 μm and the bottom electrode of 10-nm-thick epitaxial SRO layer were used for the measurement along out-of-plane direction, as shown in the inset of Figure 2(b).

**Supporting Information**
Supporting Information is available from the Wiley Online Library or from the author.

**Acknowledgements**

This research is supported by the US Department of Defense Army Research Office under grant No. W911NF0910530 and the US National Science Foundation under Grant No. DMR-1205469. The TEM facility at Florida State University is funded and supported by the Florida State University Research Foundation, National High Magnetic Field Laboratory (NSF-DMR-0654118) and the State of Florida.

Received: ((will be filled in by the editorial staff))
Revised: ((will be filled in by the editorial staff))
Published online: ((will be filled in by the editorial staff))

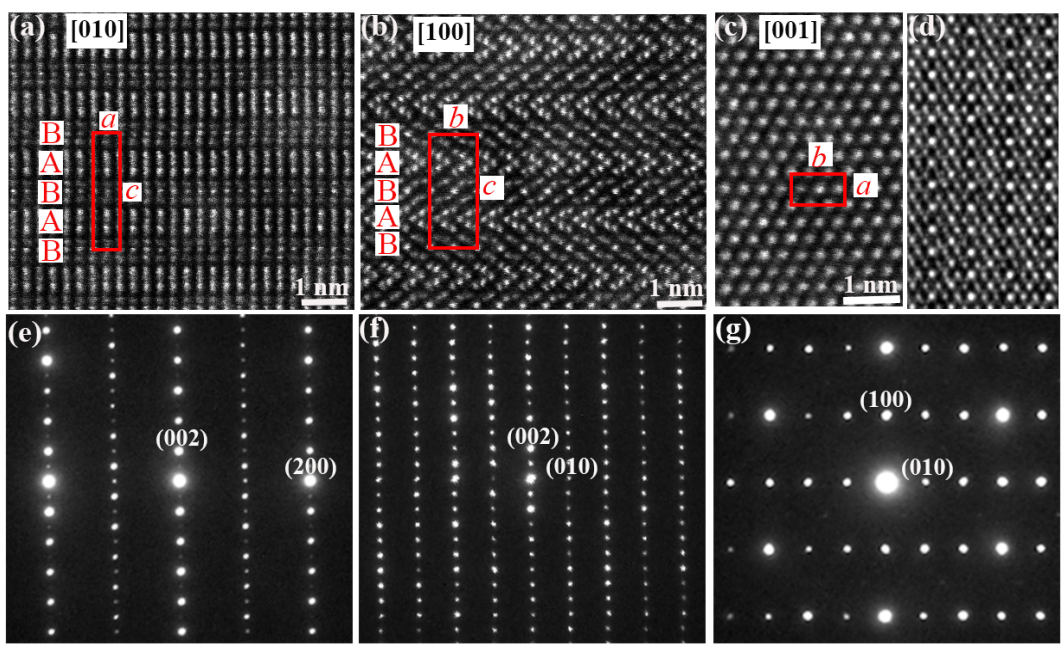

Figure 1, Miao *et al.*

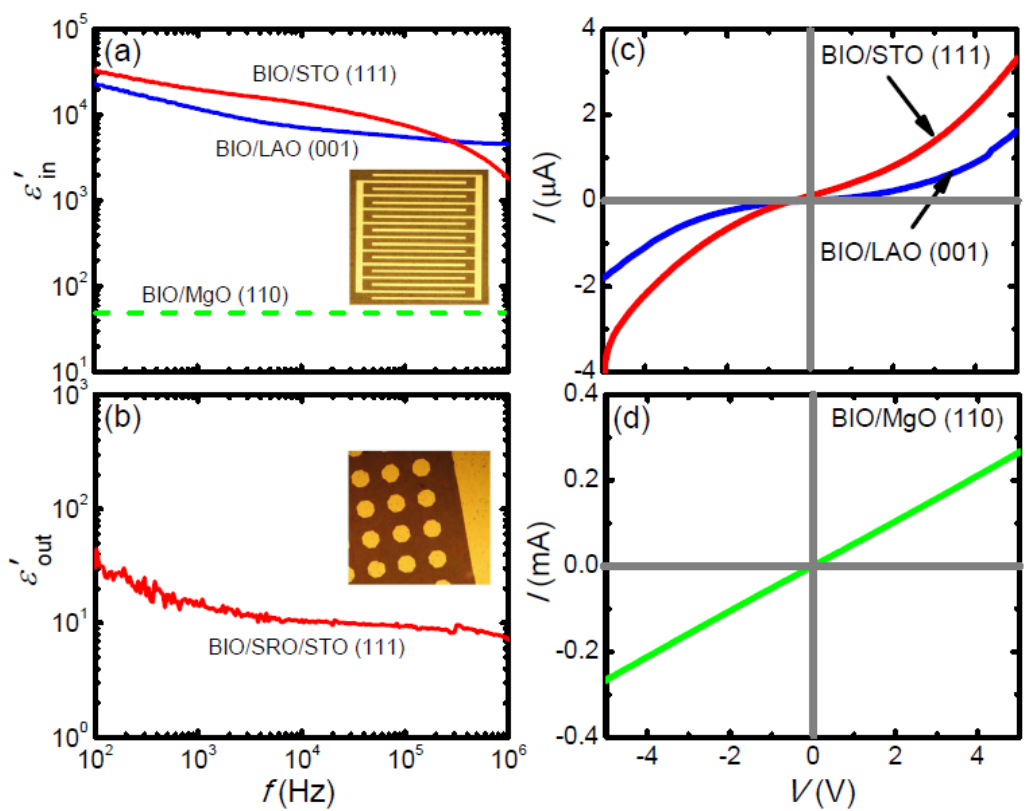

Figure 2, Miao *et al*.

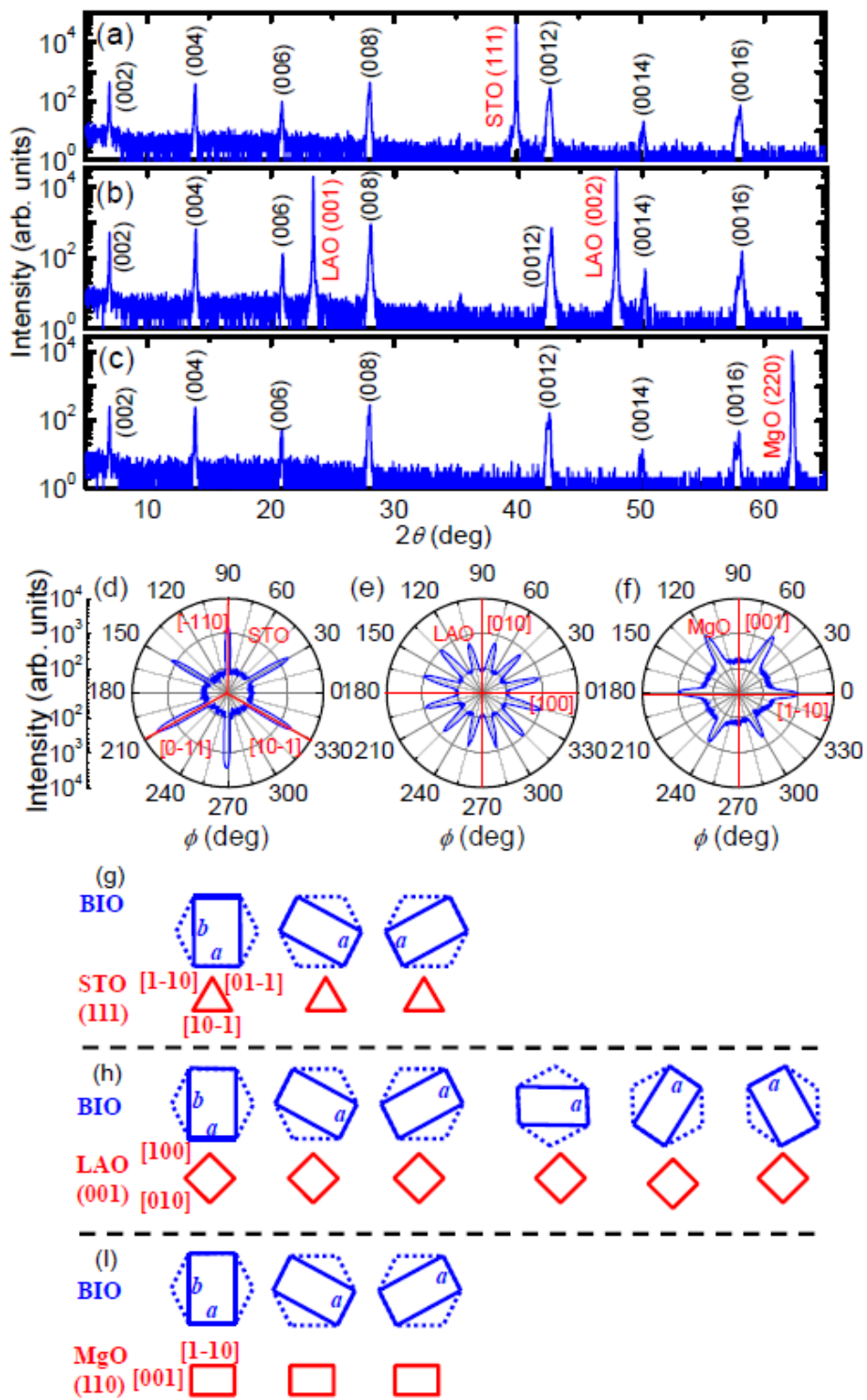

Figure 3, Miao *et al.*

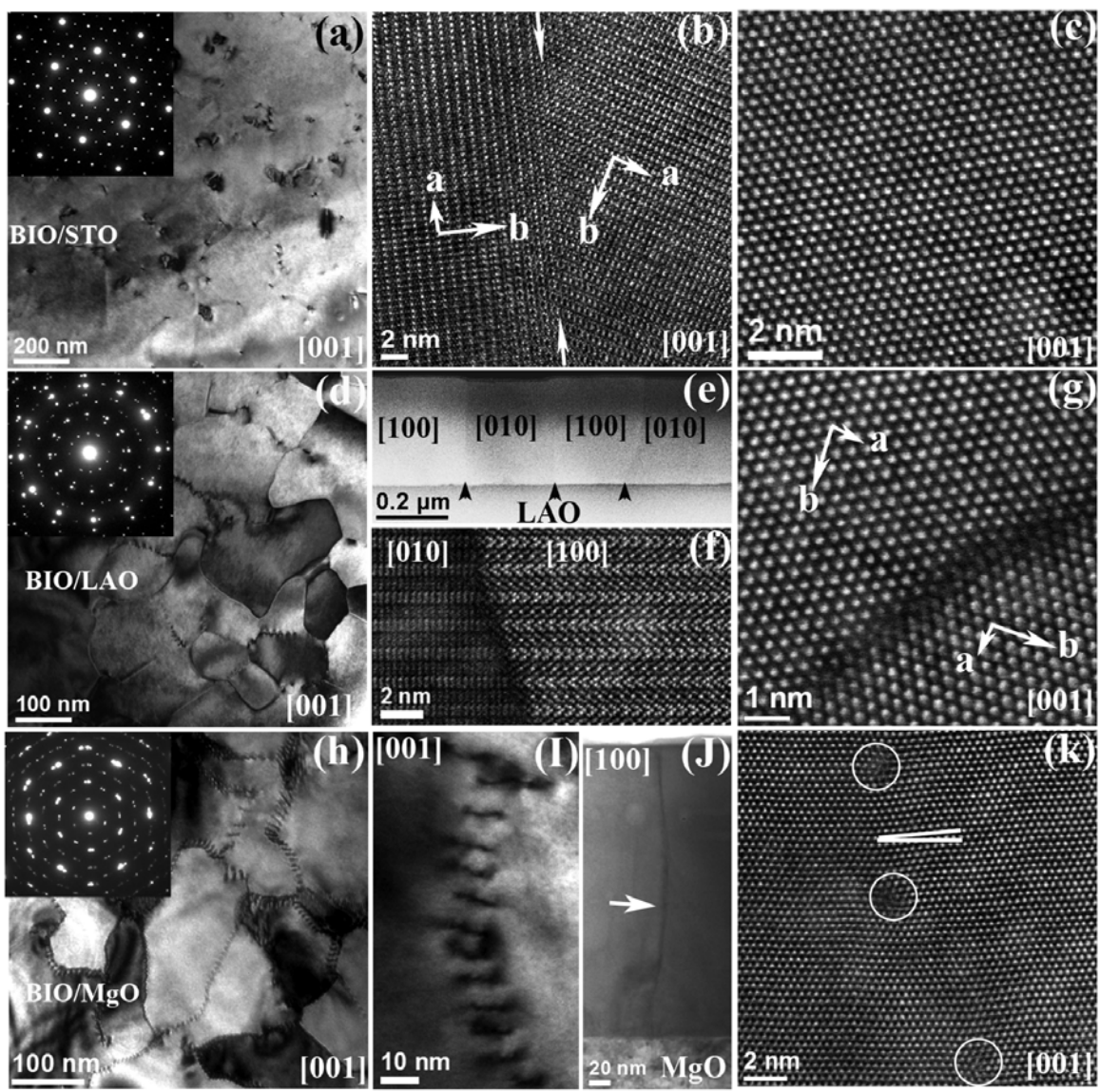

Figure 4, Miao *et al.*

**Figure 1.** Room temperature STEM HAADF atomic images and electron diffraction patterns of the BIO film. STEM HAADF images for the BIO film projected along the (a) [010], (b) [100] and (c) [001] directions. (d) HRTEM image of the film along [001]. Electron diffraction patterns taken along (e) [010], (f) [100] and (g) [001] zone axis.

**Figure 2.** Room temperature permittivity spectrum and I-V curve for the BIO films (a) In-plane permittivity $\varepsilon'_{in}$ and (b) out-of-plane permittivity $\varepsilon'_{out}$ for BIO films as functions of frequency $f$. The insets shows pictures of (a) IDE patterns for in-plane measurement and (b) top electrodes for out-of-plane measurement. In-plane I-V curve for (c) the BIO/STO(111) and BIO/LAO(001) films as well as (d) the BIO/MgO(110) film.

**Figure 3.** Room temperature HR-XRD $\theta$-$2\theta$ scan, $\varphi$ scan results and epitaxial relationships for BIO films. HR-XRD $\theta$-$2\theta$ scans of 270-nm-thick BIO films on (a) the STO(111) substrate, (b) the LAO(001) substrate and (c) the MgO(110) substrate. HR-XRD $\varphi$ scans of BIO (0 4 18) reflections for (d) the BIO/STO(111) film, (e) the BIO/LAO(001) film, and (f) BIO (2 0 16) reflections for the BIO/MgO(110) film. Schematic diagrams of epitaxial relationships for (g) the BIO/STO(111) films, (h) the BIO/LAO(001) films, and (I) the BIO/MgO(110) films. The solid lines represent the BIO and substrate lattices, and the dashed lines represent the direction of the hexagonal atomic arrangement along *ab*-plane.

**Figure 4.** Room temperature TEM/STEM images of the microstructures and electron diffraction patterns of the three films. (a)-(c) BIO/STO(111) film along [001]: (a) Low magnification BF TEM image, with inset of electron diffraction pattern. (b) HRTEM image

of type I domain boundary and (c) STEM HAADF image showing no domain bounday. (d)-(g) BIO/LAO(001) film: (d) Plan view low magnification BF TEM image, with inset of electron diffraction pattern along [001]. (e) Cross-sectional STEM HAADF image of side view of the film at low magnification showing 90° domains. (f) High magnification STEM HAADF image of the side view of the 90° domain boundary, and (g) STEM HAADF image of the 90° domain boundary along [001]. (h)-(k): BIO/MgO(110) film: (h) Plan view low magnification BF TEM image, with inset of electron diffraction pattern along [001]. (I) Enlarged BF TEM image of the domain boundary showing dislocation array. (J) BF TEM image of the cross sectional view of the film showing the dislocation threading through the whole film. (k) STEM HAADF image of the film along [001] showing the end on domain boundary dislocations (circled).

**The table of contents:**

**A novel insulating layered iridate** $Ba_7Ir_3O_{13+\delta}$ in thin film form is discovered. These films are characterized by colossal permittivity (CP) ~$10^4$ at room temperature, attributable to the colossal internal barrier layer capacitance effect at *atomically thin* domain boundaries. These findings suggest a new route to seeking novel CP materials through designing atomically thin domain boundaries.

**Keyword:**

iridates, thin films, dielectrics, colossal permittivity, domain boundaries, pulsed laser deposition

L. Miao, Y. Xin, J. Liu, H. Zhu, H. Xu, D. Talbayev, T. Stanislavchuk, A. Sirenko, V. Puli, and Z.Q. Mao*

**Title** Novel Layered Iridate $Ba_7Ir_3O_{13+\delta}$ Thin Films with Colossal Permittivity

ToC figure 55 mm broad × 50 mm high

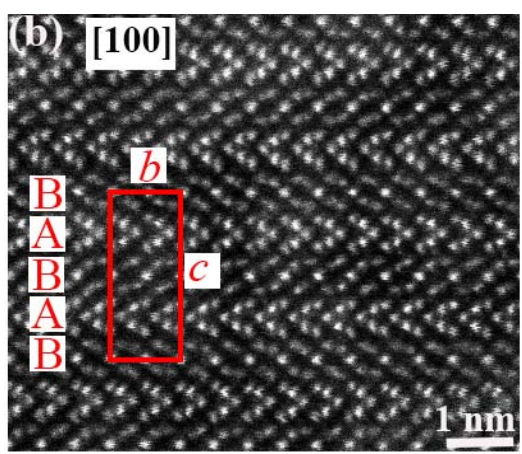

## Supporting Information

**Title** Novel Layered Iridate Ba$_7$Ir$_3$O$_{13+\delta}$ Thin Films with Colossal Permittivity

*Ludi Miao, Yan Xin, Jinyu Liu, Huiwen Zhu, Hong Xu, Diyar Talbayev, Taras Stanislavchuk, Andrei Sirenko, Venkata Puli, and Zhiqiang Mao\**

**Figure S1** shows the real part of the optical conductivity for the BIO/STO(111) film at the room temperature. We clearly see that the optical band gap of BIO is 1.3 eV at room temperature. We can get more insight into the origin of the optical conductivity features of BIO by further comparing its optical properties to the properties of the perovskite Sr$_2$IrO$_4$. The lowest energy optical transitions of Sr$_2$IrO$_4$ were interpreted as the 5$d$-5$d$ transitions between neighboring Ir ions, while the O 2$p$-Ir 5$d$ charge-transfer transitions were found at energies over 3 eV. [10,S1] The crystal-field splitting between $t_{2g}$ and $e_g$ states in the 5$d$ manifold was measured to be 2.1 eV,[S1] and the $t_{2g}$ states are further split by the relativistic spin-orbit coupling by ~0.4 eV into $J_{eff}$ = 1/2 and $J_{eff}$ = 3/2 states.[10] We tentatively assign the two broad peaks A and B at 1.8 and 2.5 eV in Fig. S1 to the transitions from the spin-orbit split $t_{2g}$ levels to the $e_g$ levels of the nearest-neighbor Ir ions. This assignment would imply that the spin-orbit splitting in BIO (0.7 eV) is higher than the splitting in the perovskite Sr$_2$IrO$_4$ (0.4 eV), a finding that needs further confirmation by both experimental and theoretical means.

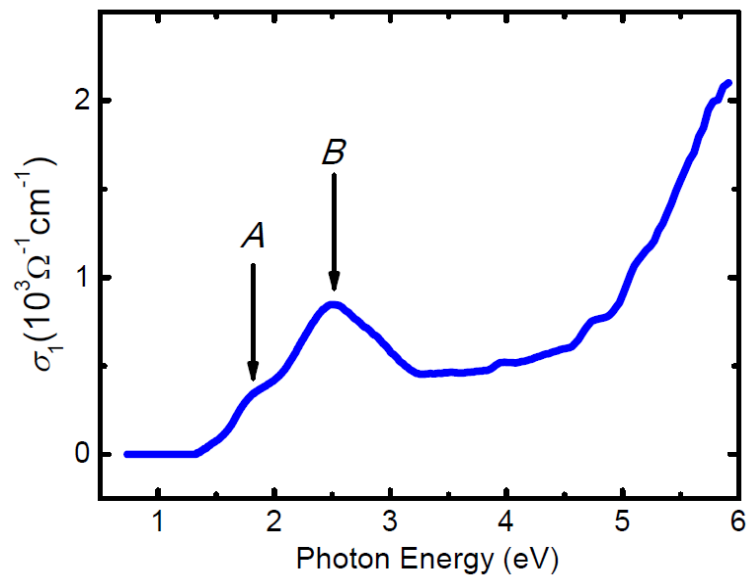

**Figure S1.** Room temperature optical spectrum for the BIO films. The real part of the optical conductivity of the BIO/STO(111) film obtained from the ellipsometry.

We have measured the tangent losses for BIO/LAO(001) and BIO/STO(111) films at room temperature from $10^2$ Hz to $10^6$ Hz, as shown in Fig. S2. The losses for both films start from a very large value of ~ $10^2$ at $10^2$ Hz, then gradually decrease to ~$10^{-1}$ at $10^6$ Hz which is 1-4 orders of magnitude larger than a typical loss value for a dielectric material with potential applications.[27] Such huge loss values are possible due to the non-empty $5d$ bands of BIO, which is distinct from $3d^0$ electronic configuration of Ti-based colossal permittivity materials such as (In + Nb) co-doped $TiO_2$[27] and $CaCu_3Ti_4O_{12}$.[28]

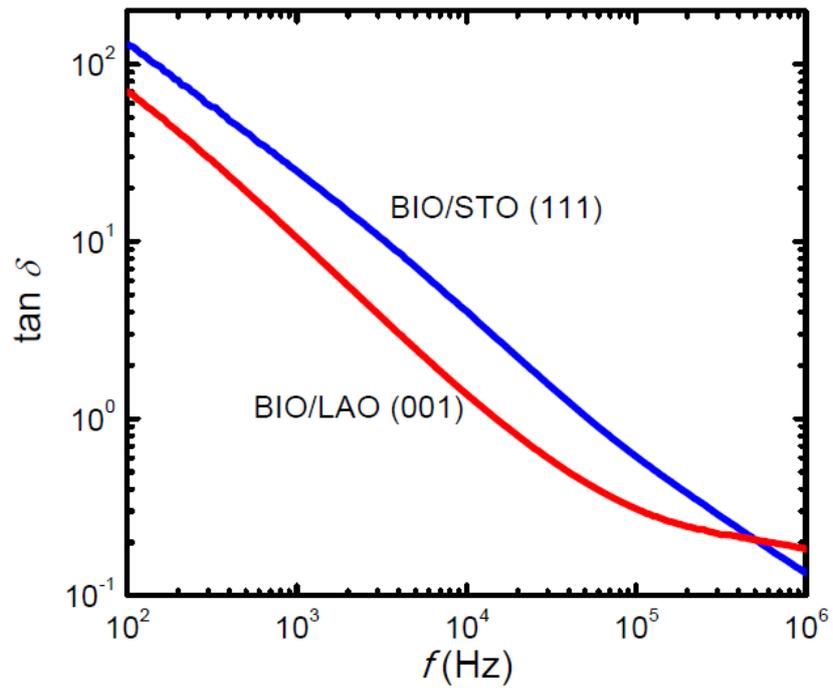

**Figure S2.** In-plane tangent losses for BIO films. In-plane tangent losses tan$\delta$ for BIO/LAO(001) and BIO/STO(111) films as functions of frequency $f$ measured at room temperature.

**Reference**

[S1] S.J. Moon, M.W. Kim, K.W. Kim, Y.S. Lee, J.-Y. Kim, J.-H. Park, B.J. Kim, S.-J. Oh, S. Nakatsuji, Y. Maeno, I. Nagai, S.I. Ikeda, G. Cao, T.W. Noh, *Phys. Rev. B* **2006**, *74*, 113104.